\documentclass[conference, 10pt]{IEEEtran}
\usepackage{cite}
\usepackage{amsmath,amssymb,amsfonts}
\usepackage{algorithmic}
\usepackage{graphicx}
\usepackage{textcomp}
\usepackage[dvipsnames*,svgnames]{xcolor}
\usepackage[hyphens]{url}

\usepackage{makecell}
\usepackage{titlesec}
\usepackage{booktabs}
\usepackage[font=normal]{caption}
\usepackage{setspace}
\pdfoutput=1
\usepackage{anyfontsize}

\def\BibTeX{{\rm B\kern-.05em{\sc i\kern-.025em b}\kern-.08em
    T\kern-.1667em\lower.7ex\hbox{E}\kern-.125emX}}
\title{\LARGE \bf LearningGroup: A Real-Time Sparse Training on FPGA via \\ Learnable Weight Grouping for Multi-Agent Reinforcement Learning}

\author{\IEEEauthorblockN {Je Yang, JaeUk Kim and Joo-Young Kim}
\IEEEauthorblockA{{School of Electrical Engineering, KAIST}\\
\texttt{\{yangje, kju5789, jooyoung1203\}@kaist.ac.kr}}
}

\begin{document}
\maketitle
\pagestyle{plain}

%%%%%% -- PAPER CONTENT STARTS-- %%%%%%%%
\begin{abstract}
Multi-agent reinforcement learning (MARL) is a powerful technology to construct interactive artificial intelligent systems in various applications such as multi-robot control and self-driving cars. Unlike supervised model or single-agent reinforcement learning, which actively exploits network pruning, it is obscure that how pruning will work in multi-agent reinforcement learning with its cooperative and interactive characteristics.
\par
In this paper, we present a real-time sparse training acceleration system named LearningGroup, which adopts network pruning on the training of MARL for the first time with an algorithm/architecture co-design approach. We create sparsity using a weight grouping algorithm and propose on-chip sparse data encoding loop (OSEL) that enables fast encoding with efficient implementation. Based on the OSEL's encoding format, LearningGroup performs efficient weight compression and computation workload allocation to multiple cores, where each core handles multiple sparse rows of the weight matrix simultaneously with vector processing units. As a result, LearningGroup system minimizes the cycle time and memory footprint for sparse data generation up to 5.72$\times$ and 6.81$\times$. Its FPGA accelerator shows 257.40-3629.48 GFLOPS throughput and 7.10-100.12 GFLOPS/W energy efficiency for various conditions in MARL, which are 7.13$\times$ higher and 12.43$\times$ more energy efficient than Nvidia Titan RTX GPU, thanks to the fully on-chip training and highly optimized dataflow/data format provided by FPGA. Most importantly, the accelerator shows speedup up to 12.52$\times$ for processing sparse data over the dense case, which is the highest among state-of-the-art sparse training accelerators.
\end{abstract}

\setstretch{0.91}
\section{Introduction}

Reinforcement learning is a promising area of machine learning, known for solving long-term decision-making problems effectively. It aims to train the action policy, which is about how an agent should take actions based on the feedback from the given environment to maximize cumulative rewards. Recently, deep reinforcement learning (DRL) that utilizes a deep neural network (DNN) as an action policy has been proposed \cite{mnih2013playing, mnih2016asynchronous, lillicrap2015continuous, fujimoto2018addressing}. Although DRL stands out in various domains such as industrial control and robotics \cite{mnih2015human, farahnakian2014energy, won2020adaptive}, all of them are limited to a single agent. Other significant applications have started to employ interaction between multiple agents, for instance, analysis of language communication and the network of self-driving cars \cite{vinyals2019grandmaster, lazaridou2016multi, palanisamy2020multi, peng2021learning}. Hence, extending DRL to have many agents is critical for developing intelligent systems where agents can interact with each other or even with people.
\par
\begin{figure}[t]
\centering
   \includegraphics[width=8.5cm, height=3cm]{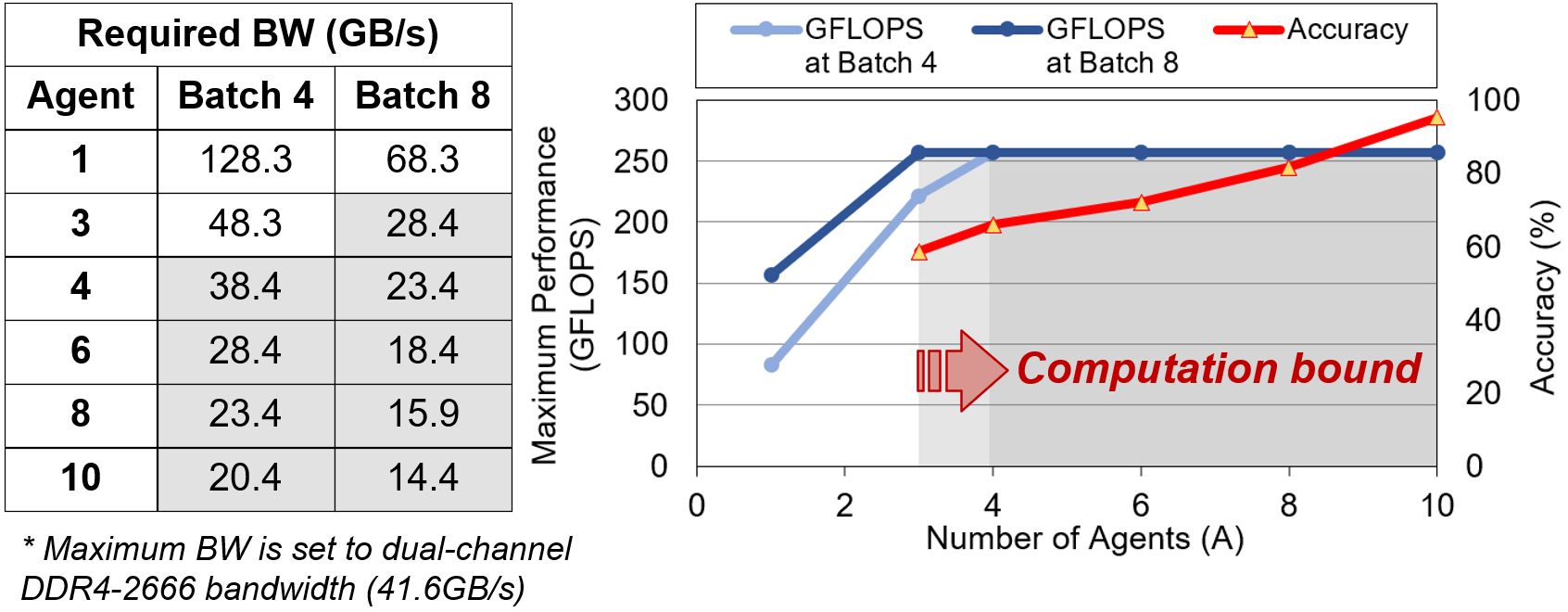}
\caption{Roofline Model of MARL in CPU System (Intel Core i5-10400 and dual-channel DDR4-2666 DIMM)}
\label{MarlCharac}
\end{figure}

There are several challenges to accelerate multi-agent reinforcement learning (MARL). First of all, training each agent independently makes a single agent's environment non-stationary, since other agents update their action policies during training. To alleviate this training instability, the latest works use a centralized network with a communication layer for flexibility, efficient exploration, and accuracy \cite{lowe2017multi,  hoshen2017vain, singh2018learning}. A long short-term memory (LSTM) layer is commonly used as communication layer to support various relationships among agents such as cooperative or competitive. Second, real-time MARL training requires high computational throughput on a small batch, which cannot be supported on conventional CPU and GPU based systems. Figure \ref{MarlCharac} shows the roofline model of MARL in current CPU system when the number of agents and batch size varies. The bandwidth requirement for a single agent case exceeds the system's maximum bandwidth, making it memory-bound. However, the system moves to compute-bound with improved MARL accuracy when the agent number increases due to the weight reuse through a centralized network. MARL requires up to 942.9 GFLOPS for effective real-time operation (8 agents, 30ms action latency), while it needs a small batch size of less than 32 for training convergence and exploration \cite{vinyals2019grandmaster}. In addition, as the MARL system is commonly used in edge environments for many applications, e.g., autonomous cars, robots, and drones, it should be power efficient as well. Current CPU and GPU-based systems cannot meet the above requirements due to the lack of computing units, high power consumption or low utilization for small batch sizes. Instead, FPGA is emerging as a new solution for a deep learning engine with real-time interaction. For example, the Xilinx U280 acceleration card provides robust computing potential through 9,024 DSPs over 41MB of on-chip BRAM while showing less power consumption than GPU. In addition, the reconfigurability of FPGA allows the optimization of irregular data access and parallelism with customized compact data format, where these hardware overhead occurs in network pruning to handle computation-bound applications.\cite{boutros2020beyond, jiang2021optimized}.
\par

In this paper, we propose a FPGA-based acceleration system named LearningGroup, to yield high performance for the real-time training of multi-agent reinforcement learning. LearningGroup successfully employs a pruning algorithm for the {\bf first} time in the on-chip training process of multi-agent reinforcement learning.
\begin{itemize}
\item {\bf Algorithm Design} LearningGroup creates computation sparsity using a weight grouping algorithm. It achieves a higher accuracy than any other types of pruning algorithms for MARL, since it flexibly masks the weights rather than removing, and trains the masks. Leveraging this flexibility, we design an accelerator that can maximally exploit the sparsity made by the pruning algorithm.
\item{\bf Architecture Design} We propose On-chip Sparse data Encoding Loop (OSEL) for efficient and hardware-amenable network pruning. It is the first sparse training accelerator that supports sparse data generation without accessing external memory. We devise a row-based load balancing scheme at run-time while taking an intra-layer parallelism together. The vector processing units achieve high utilization by operating multiple compressed rows of the weight matrix in parallel based on the load balancing.
\item{\bf System Design} We prototype our LearningGroup system on a CPU-FPGA platform by integrating the proposed algorithm and architecture design together. We successfully validate our system by running MARL algorithm called IC3Net in the OpenAI multi-agent action space.
\end{itemize}

As a result, LearningGroup with OSEL minimizes the cycle and memory space for sparse data generation up to 5.72$\times$ and 6.81$\times$, respectively. LearningGroup's FPGA accelerator achieves 7.13$\times$ speedup and 12.43$\times$ energy efficiency on average over Nvidia Titan RTX GPU without losing accuracy. Most importantly, it shows 12.52$\times$ speedup over the dense model, which is the best speedup among the state-of-the-art accelerators that handle sparse training, thanks to fully on-chip training based on algorithm/architecture co-design approach.    
\section{ Background}

\subsection{ Multi-Agent Reinforcement Learning}
\begin{figure}[t]
\centering
   \includegraphics[width=7cm, height=3.3cm]{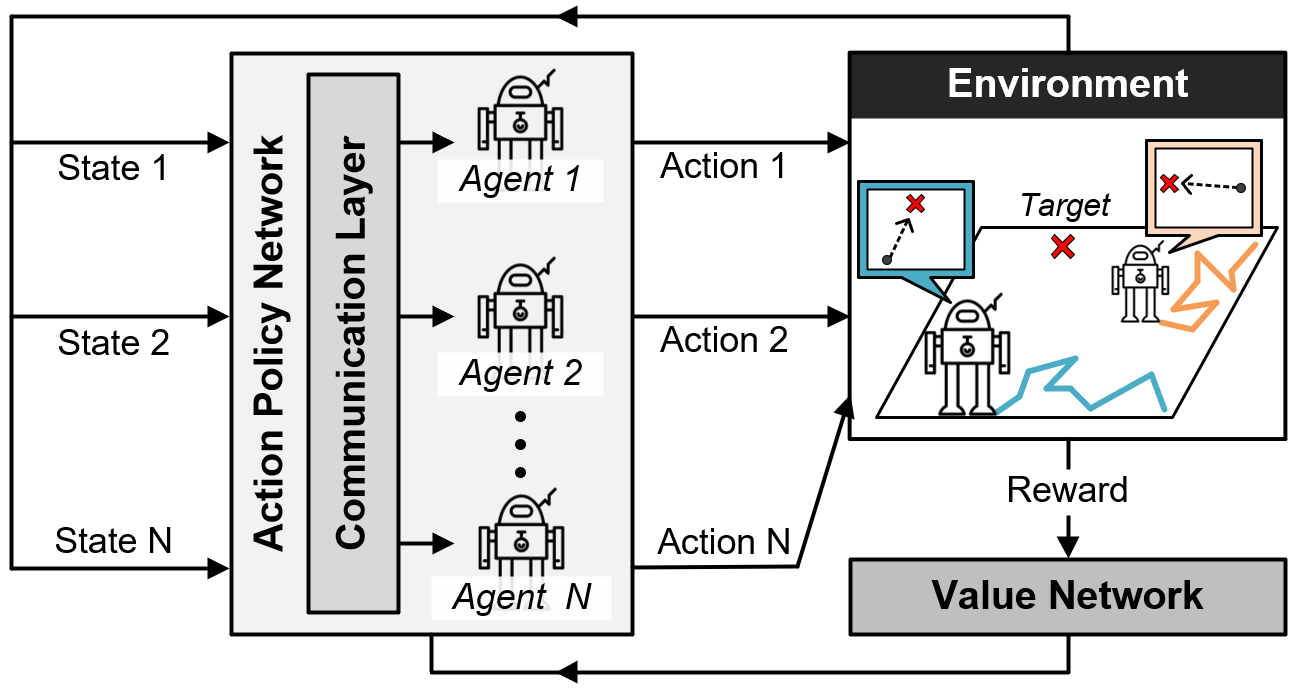}
\caption{Multi-Agent Reinforcement Learning}
\label{MARL}
\end{figure}

Recently, scaling DRL with multiple agents has played a pioneering role in addressing more complex problems such as traffic engineering, resource sharing, and data collection  \cite{vinyals2019grandmaster, lazaridou2016multi, palanisamy2020multi, peng2021learning} by showing higher accuracy than single-agent case \cite{lowe2017multi, tan1993multi}. Unfortunately, training individual agent independently makes each agent's environment non-stationary since they update their action policies concurrently. To address this training instability, multi-agent reinforcement learning algorithm employs a centralized model with communication among agents. In this model, each agent can learn the experience from others and transfer them back to increase the accuracy of the entire team. Figure \ref{MARL} depicts the overview of multi-agent reinforcement learning (MARL). In MARL, each agent receives a state from an environment considering other agents' information (i.e., states and actions) through a communication layer. Based on this encoded state, each agent takes action through the action policy network. After sending the selected actions to the environment, the agents receive a total reward value and their next states from the environment. Using the reward, the value network calculates how effective the actions and communications are. The evaluated values are passed to the action policy network to give a feedback for weight update. The agents learn action policies that maximize the reward in a given environment by repeatedly performing the above process for the newly received state. 
\par

\subsection{ Network Pruning and Sparsity Handling Accelerator}
DNN pruning takes advantage of inherent redundancy in the original dense model and removes unnecessary weights to generate a sparse model \cite{han2015deep, han2015learning}. The sparse model has less computation and memory cost with fewer parameters. In sparse model execution, pruning granularity, sparse data representation, and sparse matrix-vector multiplication (SpMV) are primary concerns for efficient processing. Many hardware accelerators have been proposed to support various types of pruning algorithms, but each type has limitations. First, element-wise pruning algorithms require sorting all the weight values or comparing them with a threshold to determine which parameters to prune. This approach is time-consuming and involves considerable irregular memory accesses in hardware due to unstructured pruning pattern \cite{lis2019full, zhang2019eager}. Second, several works gradually prune the weights with a small threshold value, but this approach suffers from a low hardware performance at the early stage due to the low starting sparsity \cite{zhang2019eager, dai2020sparsetrain}.
%Second, several works achieve a low sparsity ratio by gradually pruning the model. As the sparsity is low at the early process of training, this method cannot fully utilize the advantage of pruning. 
Lastly, most of the existing sparsity handling accelerators are designed for inference \cite{han2016eie, zhang2016cambricon, parashar2017scnn, gondimalla2019sparten, chen2019eyeriss, guo2020accelerating}, so they are not suitable for sparse training, which MARL requires to perform. Unlike sparse inference, the value of the weight and the pruning ratio are dynamically changing during the training process. In addition, as the weight data are transposed in the backward propagation, it should be reflected in the sparse data representation that includes the locations of non-zero elements. A few previous accelerators encode the sparse data on the software, not in hardware at runtime. As the sparse data continues to change in the training process, generating them off-chip (i.e., by software) would significantly increase external data movement \cite{zhang2019eager, dai2020sparsetrain, yang2020procrustes}.
\par
In this paper, our motivation is to build an efficient sparse training system for compute-bound MARL. To this end, we identify an optimized pruning algorithm and design the hardware based on it, which can generate and process sparse data on-chip. Putting together the requirements for sparse MARL training, such as reconfigurability for handling irregular data pattern, high computation potential, and low power consumption, we decide to integrate the system on FPGA.
\section{ LearningGroup System Design}
\begin{figure}[t!]
\centering
   \includegraphics[width=8cm, height=5.5cm]{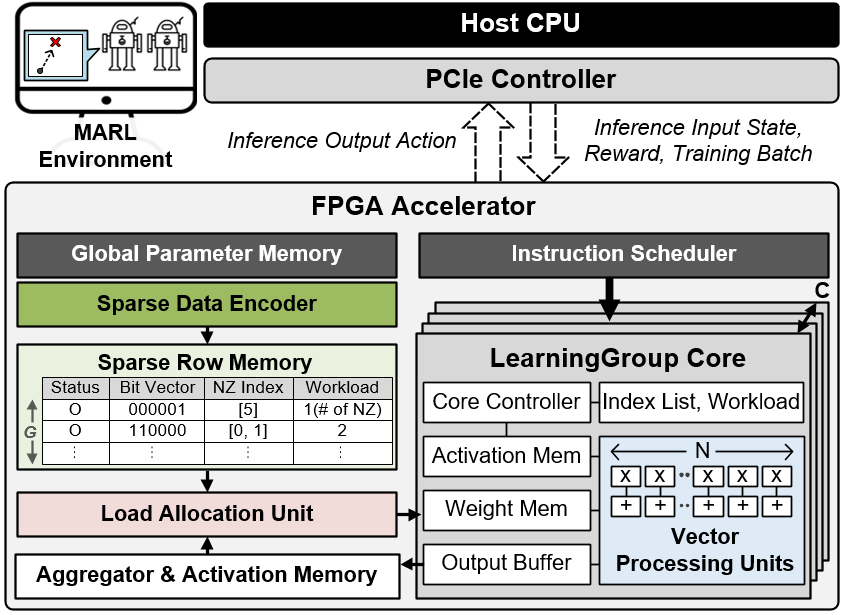}
\caption{LearningGroup's System Architecture}
\label{Overall Architecture}
\end{figure}

Figure \ref{Overall Architecture} shows the architecture of the proposed LearningGroup system. The host CPU emulates the reinforcement learning environment and the FPGA accelerator runs compute-intensive DNNs of multi-agent reinforcement learning. The host initiates the FPGA accelerator by giving a DNN training dataset and system configuration through the PCI express, and the accelerator gives the action as an output to the host and updates its model parameters.
\par
In the FPGA accelerator, the instruction scheduler is the main control unit that runs the four major operational stages: weight grouping, forward propagation, backward propagation, and weight update. In the weight grouping stage, the sparse data encoder generates sparsity in weights based on an optimized weight grouping algorithm and stores the sparse data to the sparse row memory. Then, the load allocation unit accesses the global parameter memory that stores all the model parameters by referring the sparse row memory. It reads only unmasked weights to skip unnecessary computations and performs load balancing when it distributes the activations and weights to the cores. For the forward and backward propagation, the scalable $C$ number of LearningGroup cores perform matrix/vector operations in parallel. Their vector processing units are customized to process the irregular non-zero data received from the load allocation unit. The aggregator combines the partial sums from each core and sends the final result to the load allocation unit for the workload assignment of next layer. The above process is repeated for the entire layers of the policy and value network. Finally, the weights are updated with the propagated errors and calculated gradients, then the weight grouping is performed again at the new iteration with respect to the updated weights.

\subsection{ Pruning Algorithm Selection} 
\begin{figure}[t!]
\centering
   \includegraphics[width=8cm]{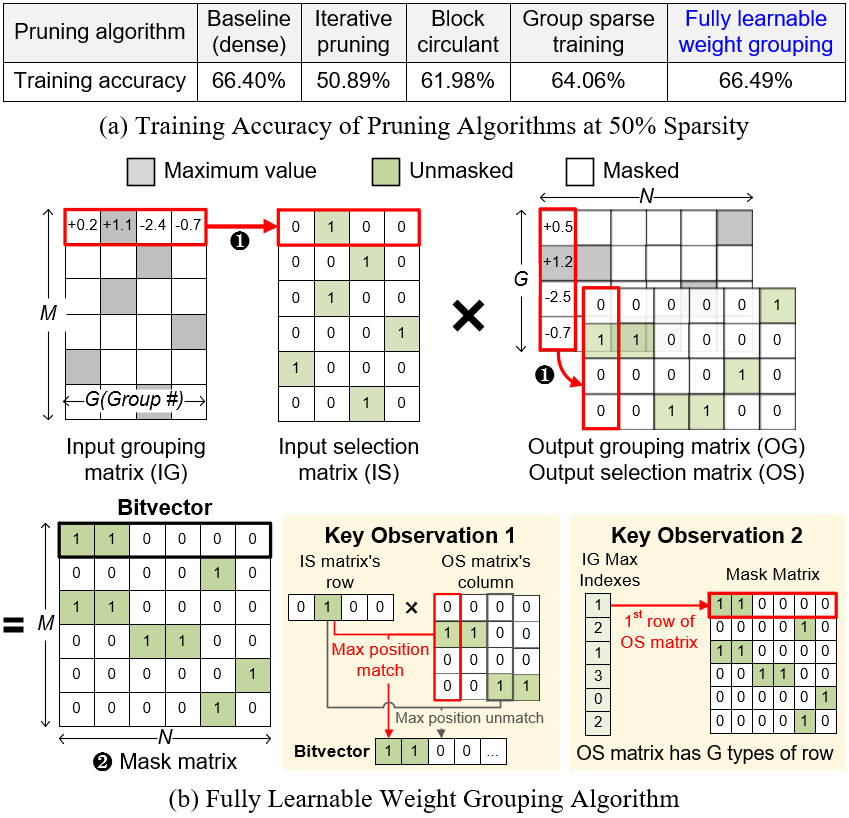}
\caption{Fully Learnable Weight Grouping Algorithm (FLGW)}
\label{FLGC}
\end{figure}

The pruning algorithm is a key component in the LearningGroup system for real-time and energy-efficient on-chip inference and training. It reduces both computational load and memory access by utilizing only effective parts of the model with negligible accuracy loss. We evaluate the state-of-the-art pruning algorithms studied in CNNs to investigate which is suitable for MARL. We choose four candidates: iterative pruning, block circulant, group sparse training, and fully learnable weight grouping algorithm. Iterative pruning \cite{zhang2019eager, yang2020procrustes, dai2020sparsetrain} eliminates the parameters with the smallest value every iteration, so the pruning ratio increases as the training progresses. It is not hardware-friendly because it is time-consuming to sort the parameter values every iteration. Block circulant method \cite{narang2017block} compresses the network parameters in a block format. This structured pruning has an advantage in encoding cost, but it suffers from a low compression ratio. Group sparse training \cite{lee2021gst} has been proposed to overcome the disadvantage of the block circulant method. It utilizes the block circulant compression and selectively adopts iterative pruning within the block to reach a target sparsity. As it already compresses parameters using the block circulant method, pruning even more parameters within the block may negatively affect the behaviors of agents in MARL, in which all agents share a centralized network. Lastly, fully learnable weight grouping (FLGW) \cite{wang2019fully} creates data sparsity using grouping matrices. It introduces an input and output grouping matrix in each layer and generates a mask matrix based on the grouping matrices for pruning. In this approach, the input and output grouping matrix are being trained independently from the training of the original data.
Figure \ref{FLGC}(a) shows the training accuracy of the selected pruning algorithms when they are applied to a multi-agent reinforcement learning network named IC3Net\cite{singh2018learning}. The baseline case that uses the dense model without any pruning shows 66.4\% accuracy. The FLGW achieves the highest accuracy among the other pruning algorithms by having separate and learnable matrices for pruning rather than removing actual parameters of the network. Thus, we decide to adopt the fully learnable weight grouping as the pruning algorithm for our LearningGroup system.
%for MARL and implement it in a hardware-amenable way. 

Figure \ref{FLGC}(b) illustrates how the fully learnable weight grouping algorithm creates sparsity in detail. For a layer with the size of ${M \times N}$ that converts ${1 \times M}$ input vector to ${1 \times N}$ output vector, it creates the input grouping (IG) and output grouping (OG) matrix whose sizes are set to ${M \times G}$ and ${G \times N}$, respectively, where $G$ is the number of groups. Both grouping matrices are initialized randomly. First, it finds the maximum value for each row of the IG matrix and binarizes each row by assigning 1 to the maximum location and 0 to the rest to generate the input selection (IS) matrix (red box in the figure). Likewise, it finds the maximum value for each column of the OG matrix and generates the output selection (OS) matrix. Finally, the mask matrix, whose size is the same as the layer size ${M \times N}$, is generated by multiplying the IS matrix and OS matrix. The FLGW uses an indirect approach in pruning: it does not remove the weight values but exclude some weights by referring to the mask matrix. In other words, the weights with the value '1' in the mask matrix survive and the sparsity level can be adjusted using the number of groups as the FLGW selects only one element among $G$ values.
\par
The FLGW provides more flexibility than the other pruning methods because it trains the grouping matrices and determines what to mask for each iteration. It preserves the original weight values as the masked weights can be used in the next iteration by not setting them to zero. Meanwhile, the values of each grouping matrix are trained based on the errors of the corresponding selection matrix and the mask matrix is newly generated every iteration with the update of grouping matrices. Most importantly, the weight grouping algorithm can keep the model accuracy because its model pruning through the mask matrix shows the pattern of unstructured pruning. Leveraging this flexibility, we propose an efficient sparse data generation and sparse matrix-vector multiplication in hardware.
\subsection{ On-chip Sparse Data Generation for Weight Grouping} \label{section OSEL}
To the best of our knowledge, LearningGroup is the first real-time learning system that generates sparse data on-chip for both inference and training. For efficient mask generation in hardware, we propose a sparse data generation scheme named On-chip Sparse data Encoding Loop (OSEL) with the two key observations (Figure \ref{FLGC}(b)). We use the term \textit{bitvector} to refer to each row of the mask matrix. {\bf Our first observation is that an element of a bitvector is 1 if and only if the maximum positions of corresponding IS matrix's row and OS matrix's column are the same.} Since only one element of IS matrix's row and one element of OS matrix's column are 1, their inner-product is 1 only if their maximum indexes are same. {\bf Our second observation is that the possible number of bitvector values in the mask matrix is bound to the group number $G$.} As the mask matrix is generated by multiplying the IS and OS matrix, the resulting mask matrix consists of the rows of OS matrix. To be more specific, if the $i^{th}$ column is 1 in the IS matrix's row, the resulting bitvector of the mask matrix is same as the $i^{th}$ row of the OS matrix. For example, if the first row of IS matrix has 1 in the second column, i.e., \texttt{0100}, the first row of the mask matrix will be the same as the second row of OS matrix, i.e., \texttt{110000}. Therefore, the rows of the mask matrix are made of the rows of OS matrix, whose possible maximum number is $G$.
\par

\begin{figure}[t!]
\centering
   \includegraphics[width=7.5cm]{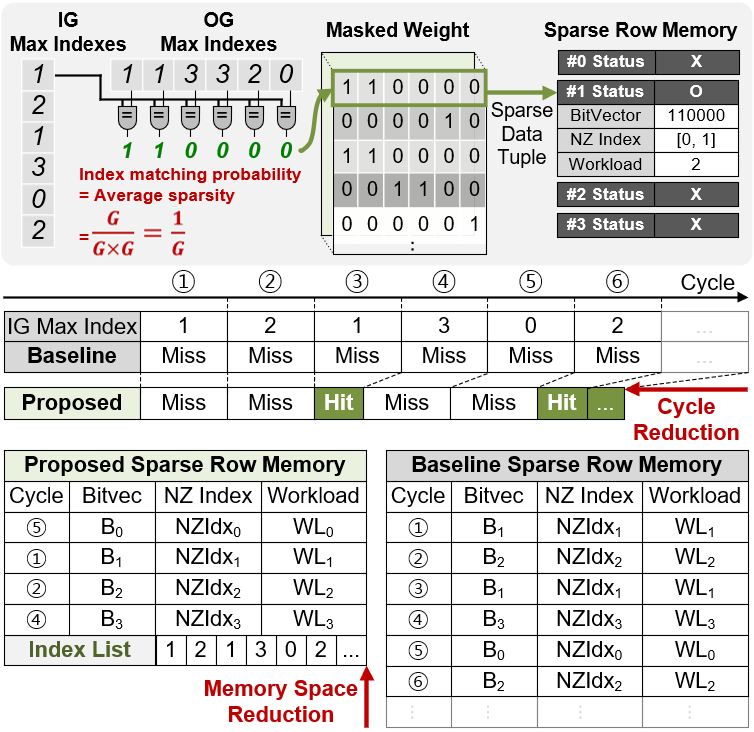}
\caption{OSEL Implementation in Sparse Data Encoder}
\label{OSEL}
\end{figure}

Using the above two observations, Figure \ref{OSEL} shows how OSEL is implemented in the sparse data encoder with an example sequence of the group number 4. The sparse data encoder receives two index lists as input: the list that has the maximum index of each row of IG matrix and the list that has the maximum index of each column of OG matrix. For output, it includes the sparse row memory that stores a custom tuple information (a bitvector, non-zero indexes, and a workload) for each row of the mask matrix. The non-zero indexes mean the locations of the unmasked weights in the row and the workload means the number of the unmasked weights. 
The sparse data encoder receives the maximum index value of the IG matrix's row every cycle. Then, it checks the status of max index's row in the sparse row memory to see if the sparse data tuple has been generated. If not (\texttt{status=X}), it needs to calculate the bitvector (\texttt{Max Index Miss}). Using our first observation, OSEL simply compares the max index from the IG matrix against the max indexes from the OG matrix in parallel to generate the bitvector. Once OSEL finishes the comparison, it stores the bitvector, non-zero indexes, and workload in the sparse row memory and updates the status. It also stores the max index in the index list to refer to the sparse data tuple that corresponds to the row of weight matrix. On the other hand, if the bitvector has been generated before (\texttt{status=O}), it just stores the max index to the index list then skips sparse data generation (\texttt{Max Index Hit}).
The bottom part of Figure \ref{OSEL} describes how LearningGroup generates the entire sparse data using OSEL cycle by cycle. At cycle 1 and 2, the bitvectors for the index 1 and 2 have not been generated yet in the sparse row memory, so OSEL newly generates the bitvector using comparators and stores the sparse data tuple. At cycle 3, as the max index 1 is found in the sparse row memory, the sparse data encoder does not update the tuple but stores the max index in the index list. At cycle 4 and 5, the sparse row memory update occurs for the index 3 and 0, respectively. At this point, the sparse row memory stores all the possible bitvectors for $G$ different rows, making a complete mask matrix. Thus, starting from cycle 6, the sparse data encoder always hits the index from the sparse row memory. 
\par
Comparing with the original weight grouping algorithm, the sparse data encoder only compares the index rather than expensive matrix multiplication to generate the bitvector thanks to the first observation. Moreover, our second observation enables LearningGroup to save both cycles and on-chip memory space using row-wise information caching. For the baseline case without OSEL's caching, sparse data encoder always calculates the bitvector and stores a sparse data tuple for every maximum index. With OSEL, it stores only the essential data to the on-chip memory and replaces redundant calculation and memory footprint by referencing the on-chip memory with the index list. In addition, OSEL can be applied to the training process with simple modification. As the backward propagation uses transposed matrices, it regards OG matrix as IG matrix and compares the max index from the OG matrix against the max indexes from the IG matrix one by one to generate a bitvector. Once a bitvector for a row of the transposed matrix is generated, it updates the sparse row memory with sparse data tuple same as the inference. The sparse data tuple generation for training can be operated in parallel with the inference computation so its latency can be hidden.

\begin{figure}[t!]
\centering
   \includegraphics[width=7.8cm, height=6.5cm]{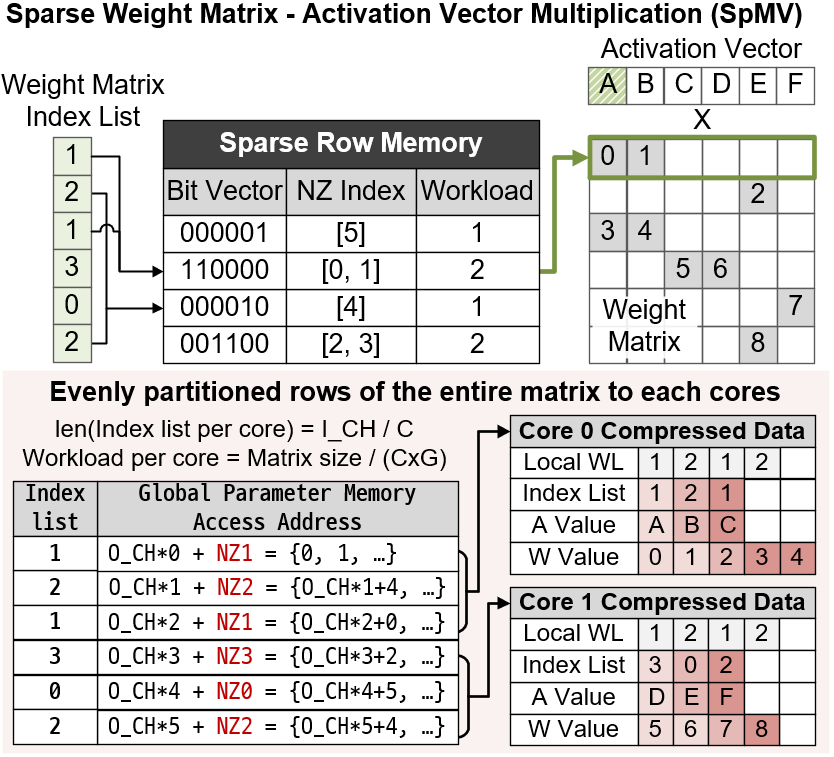}
\caption{Memory Management and Workload Allocation}
\label{compress allocation unit}
\end{figure}

\subsection{ Memory Management and Load Balancing for Parallelized On-chip Sparse Training}
Although unstructured pruning like bitvector is preferable for training accuracy, its non-uniformly distributed zeros cause load imbalance problem with irregular memory access pattern, which seriously harm the hardware utilization as well as performance in a sparse training accelerator. Some of the previous accelerators utilize software-based load balancing techniques that do not require any additional hardware cost. However, software-based load balancing is only available to the static sparsity, whose sparse locations are already known in compile time and do not change at run-time.

Figure \ref{compress allocation unit} describes how the load allocation unit accesses the global parameter memory and assigns fetched weights with activations to each core. It performs load balancing when it sends activation and weight to each core by referring to the sparse row memory. It receives the index list from the sparse data encoder, which stores the row indexes of the weight matrix that have to be accessed in order. Since the sparse data encoder has already calculated the positions of unmasked weights for each row (i.e., non-zero indexes), the load allocation unit simply refers to the sparse row memory and uses them for the address calculation of the global parameter memory. Depending on the number of rows, the output channel of the weight matrix is used as offset, and then added with the non-zero index to obtain the memory address where the unmasked data is stored. The memory address of transposed weight can be calculated by referencing to the sparse row memory for training and using input channel as offset. 
\par
LearningGroup has to balance the computational workload among cores in the hardware at run-time, since it dynamically generates the sparse data and updates it every iteration. To this end, the load allocation unit evenly partitions the rows of entire matrix by the number of cores based on the first observation described in Section \ref{section OSEL}. Since the maximum index of each row of the IG matrix and the maximum index of each column of the OG matrix must match to set the bitvector 1, the probability for that is $1/G$ and it can be interpreted as the average sparsity. Therefore, if we evenly distribute the rows to the cores, the workload of each core would be converged into $1/(C \times G)$ of the total workload, where $C$ is the number of cores. For example, the load allocation unit evenly allocates three weight matrix rows and activations to each core, as shown in the bottom part of Figure \ref{compress allocation unit}. It also distributes the weight matrix index list according to each matrix row. Since LearningGroup already adopts the row-wise computing, the row-based load balancing scheme does not require any additional logic such as counter or shifting. In addition, as it distributes the workload out of a single layer into multiple cores, we can further utilize the intra-layer parallelism.

\begin{figure}[t!]
\centering
   \includegraphics[width=7.8cm]{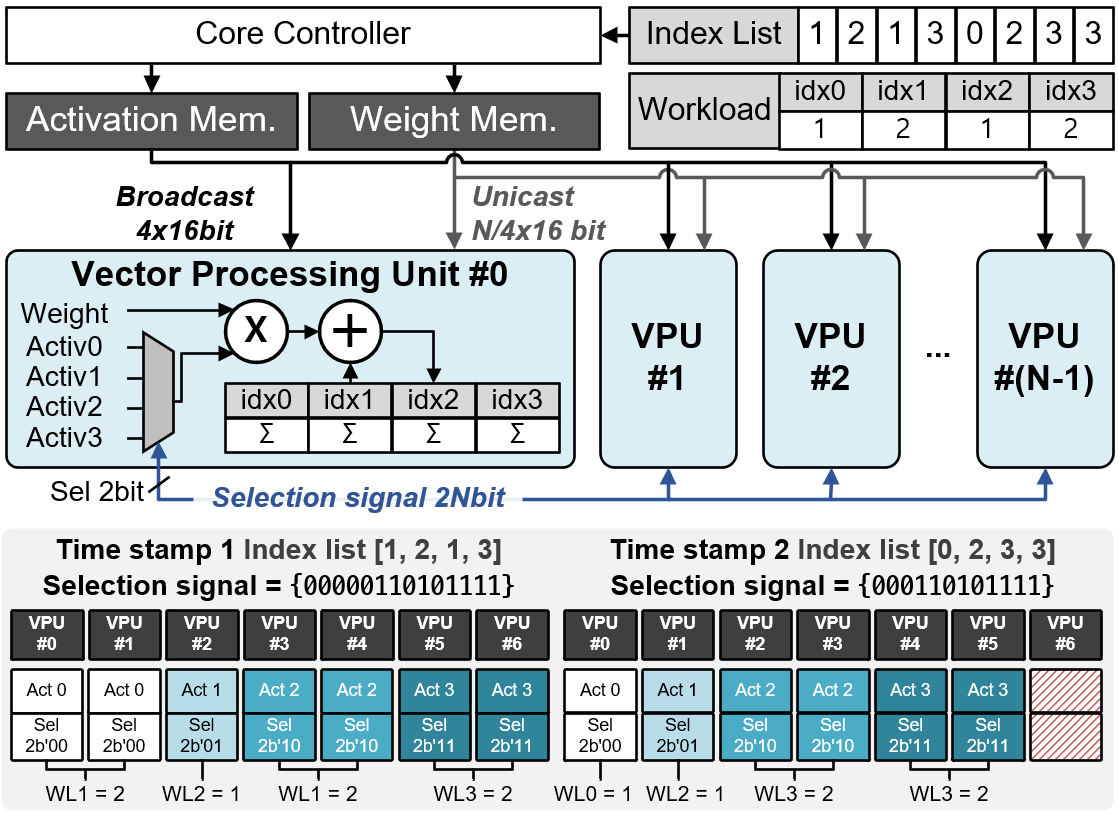}
\caption{LearningGroup Core's Vector Processing Unit Dataflow}
\label{LearningGroup Core}
\end{figure}

\subsection{ Dense/Sparse Vector Processing Unit}
Figure \ref{LearningGroup Core} shows the architecture of LearningGroup core that performs matrix-vector multiplication of DNN for both dense and sparse case and sends the result to the aggregator. It broadly consists of the core controller, activation memory, weight memory, index list with index-specific workload, and the $N$ number of dense/sparse vector processing units (VPUs). The activation and weight memory stores the compressed data from the load allocation unit. By flattening the workloads into one-dimensional vector, the LearningGroup core processes up to four rows of the weight matrix simultaneously, in which each row has a different workload. The core controller broadcasts four activations to the VPUs while it loads the individual weights during 4 cycles. To select the proper activation among the four broadcasted activations, each VPU uses 2-bit selection signal. The controller generates an array of selection signals for VPUs based on the index list and workload values for four rows: the $WL_0$ number of VPUs select $Activation 0$, the $WL_1$ number of VPUs select $Activation 1$, and so on. For example, the controller reads a list of four indexes \texttt{[1, 2, 1, 3]} at time stamp 1. Then, it reads the workload corresponding to each index and makes a selection signal according to the number of workloads. Even if the index list changes at time stamp 2 to \texttt{[0, 2, 3, 3]}, the number of selection signals can be changed easily using pre-calculated workload so that the operation can be performed compactly.
\par
Each VPU includes a FP16 multiplier, a FP16 adder, and a 4-to-1 multiplexer. It chooses an activation value from a 4-to-1 multiplexer with the given selection signal. It also includes four independent accumulation registers, so each register can accumulate the partial sums for the corresponding row index. We choose 264 for the number of VPUs ($N$) considering the network dimension and the shift amount of the select signal generation. With this configuration, the LearningGroup core shows high compute utilization of 86.96\% and 96.89\% on average for the dense and sparse MAC operations, respectively.
\section{ Evaluation} 
\subsection{ Methodology} \label{sec:eval-methodology}
{\textbf{Algorithm Validation}}
To evaluate the LearningGroup system, we run physical locomotion benchmark named \textit{Predator-Prey-v2} from MuJoCo physics engine \cite{mordatch2017emergence}. In \textit{Predator-Prey-v2}, we have \textit{A} number of cooperative agents trying to find a stationary prey. Each time the cooperative agents collide with a prey, the agents are rewarded. Individual agents observe their relative positions and velocities. We choose IC3Net \cite{singh2018learning} algorithm for multi-agent reinforcement learning and use the same network configuration as in the original work. The network parameters are optimized by the RMSprop \cite{tieleman2012lecture} optimizer with a learning rate of 0.001. We set the minibatch size $B$ for each iteration (i.e., a weight update per $B$ batch) and run the task for 2000 iterations in total. We measure the number of successes in catching prey every 50 timesteps, and calculate the average success rate (\%) as an accuracy.

\begin{figure}[t!]
\centering
   \includegraphics[width=8cm]{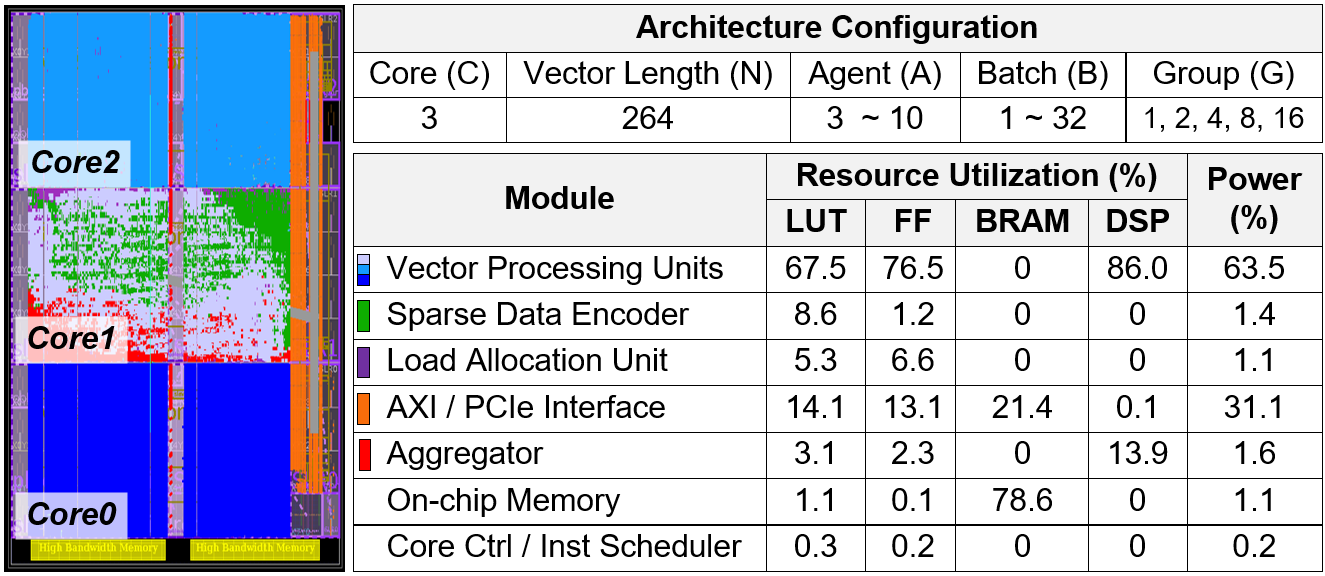}
\caption{LearningGroup Accelerator Layout and Resource Utilization on FPGA}
\label{Architecture config}
\end{figure}

{\textbf{Hardware Implementation}}
Our system is prototyped on a Xilinx Alveo U280 acceleration card using Vitis framework 2020.1 for PCIe communication. It runs at 175MHz with half-precision floating-point data. Our system has a scalable design with a variable number of LearningGroup cores, making it easy to change the level of intra-layer parallelism and prototype on modern FPGAs that have multiple super logic regions (SLRs). Considering the resource of each SLR and our core while minimizing the SLR crossing delay, we implement one LearningGroup core per SLR. As a result, we successfully integrate three LearningGroup cores with a sparse data encoder and load allocation unit on the  Figure \ref{Architecture config} shows the final layout of the FPGA and its resource utilization with architectural configuration. We design the accelerator to support the grouping up to 16, considering the trade-off between algorithmic accuracy and sparsity. It also supports various agent numbers as well as batch sizes. The vector processing units of the LearningGroup cores occupy a majority of LUT, FF, and DSP resources as well as power consumption with lots of floating-point arithmetic units. The sparse data encoder, a key enabler of the weight grouping based pruning algorithm, takes only 8.6\% of LUTs, 1.2\% of FFs, and 1.4\% of power which implies that our system effectively supports sparsity handling with minor hardware overhead.

%We integrate a sparse data encoder, load allocation unit, and three LearningGroup cores on the FPGA. Figure \ref{Architecture config} shows the final layout of the FPGA and its resource utilization with architectural configuration. We design the accelerator to support the grouping up to 16, considering the trade-off between algorithmic accuracy and sparsity. It also supports various agent numbers as well as batch sizes. The vector processing units of the LearningGroup cores occupy a majority of LUT, FF, and DSP resources as well as power consumption with lots of floating-point arithmetic units. The sparse data encoder, a key enabler of the weight grouping based pruning algorithm, takes only 8.6\% of LUTs, 1.2\% of FFs, and 1.4\% of power which implies that our system effectively supports sparsity handling with minor hardware overhead.

\subsection{ Algorithmic Accuracy}
\begin{figure}[t!]
\centering
   \includegraphics[width=8.3cm, height=2.8cm]{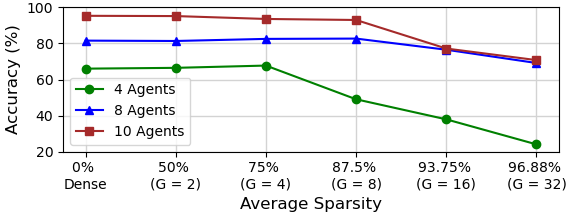}
\caption{Training Accuracy by Varying Sparsity}
\vspace{-0.1in}
\label{Algorithm accuracy}
\end{figure}

To evaluate the algorithmic accuracy of the proposed LearningGroup system, we measure the total accuracy achieved during the training process of 2000 iterations. Figure \ref{Algorithm accuracy} shows the training accuracy results for 4, 8, and 10 agents when the group number $G$ varies among 1, 2, 4, 8, 16, and 32. The average sparsity for each group number is 0\%, 50\%, 75\%, 87.5\%, 93.75\%, and 96.88\%, respectively. As it is a cooperative setting in \textit{Predator-Prey-v2}, the systems with more agents achieve higher accuracy. The accuracy tends to drop from the dense case as the sparsity increases for all cases. However, it is noteworthy that the accuracy remains as high as the dense case until the group number of 4, whose sparsity is remarkably high as 75\%. In addition, the result shows that the grouping is more favorable for many agents; the network with 8 or 10 agents maintains the accuracy even at $G=8$. With this experiment, we confirm that the weight grouping is a viable pruning solution for the MARL's centralized network.

\begin{figure}[t!]
\centering
   \includegraphics[width=8.5cm]{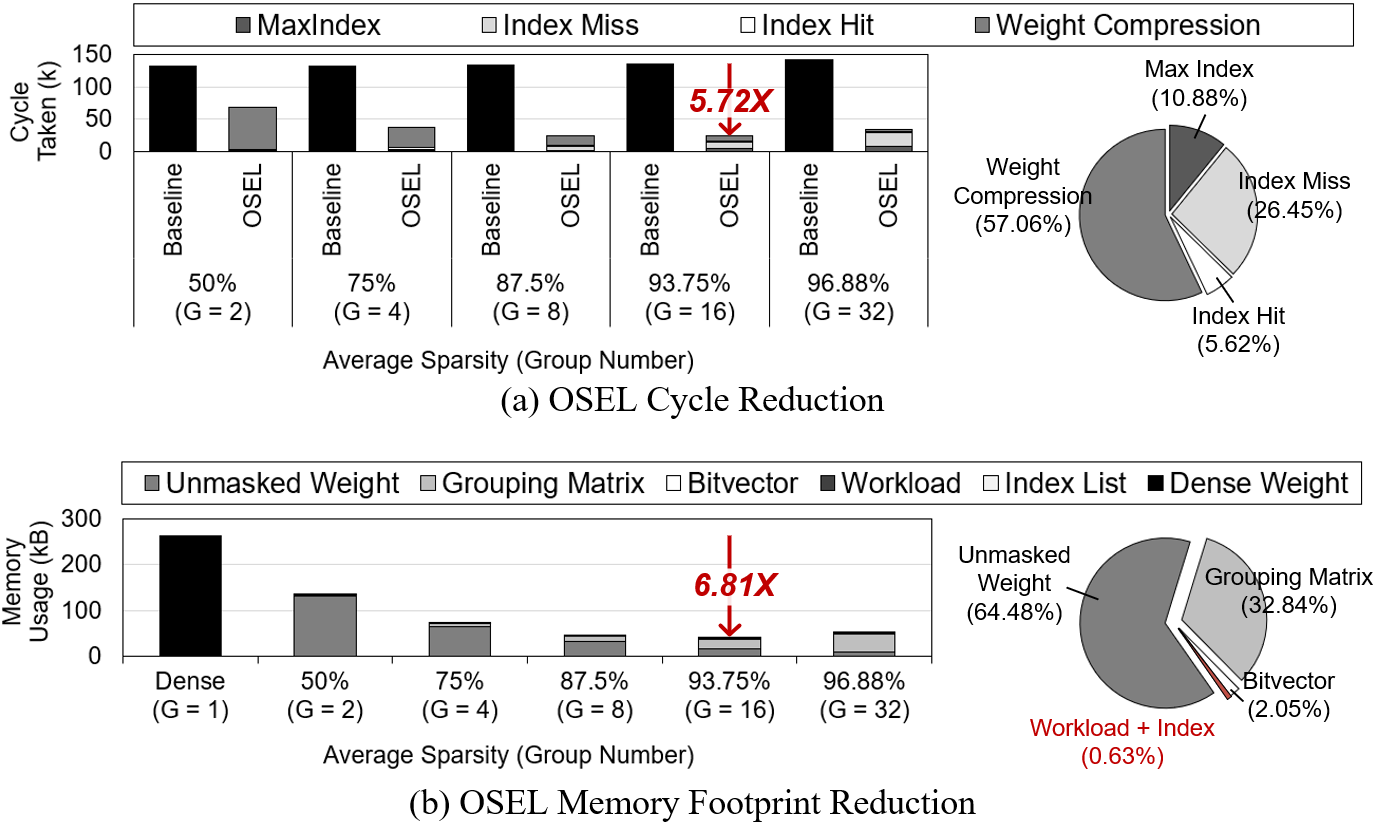}
\caption{OSEL Sparse Data Generation Efficiency}
\label{OSEL result}
\end{figure}

\subsection{ On-chip Sparse Data Generation Efficiency}
To evaluate the effectiveness of the proposed OSEL's sparse data generation, we compare the OSEL's cycle count and memory footprint with the baseline's when generating a mask matrix of size $128\times512$ with varying $G$ among 2, 4, 8, 16, and 32. The baseline system generates the mask matrix every iteration by comparing max indexes of the IS's rows with max indexes of the OS's columns. On the other hand, the LearningGroup with OSEL replaces the redundant index comparisons with bitvector caching.
\par
Figure \ref{OSEL result}(a) depicts the cycle taken for on-chip sparse data generation and compression. For the baseline system, the cycle count increases with the group number $G$ because it takes more time to find the max index from each row and column as a large $G$ makes large group matrices. Conversely, the cycle count of OSEL decreases until $G$ is less than 32, since it eliminates the unnecessary bitvector generation time. In OSEL, the execution time can be further divided into four categories. \texttt{MaxIndex} is the time for finding maximum indexes out of the grouping matrices and \texttt{Index Miss} is the time for calculating bitvectors and storing the tuple information in the sparse row memory when the index is missed. \texttt{Weight Compression}, which is the time for accessing unmasked weights, linearly decreases with the group number because the network gets sparser. Since \texttt{MaxIndex} and \texttt{Index Miss} increase with the group number, they eventually become the bottleneck at $G=32$. The pie chart on the right shows the average cycle count breakdown. As a result, OSEL accelerates the sparse data generation process by up to 5.72$\times$.
\par
Figure \ref{OSEL result}(b) shows the reduction of the sparse data memory footprint and its breakdown which only accounts for the parameters used in the actual operation (i.e., unmasked weight). The leftmost bar represents a size of dense matrix without weight grouping. As the group number $G$ increases, the memory space of unmasked weights linearly decreases in the LearningGroup system. On the other hand, the memory space of the grouping matrices increases with the group number because one dimension of each grouping matrix is $G$. Since the memory reduction effect by storing only unmasked weights is dominant, the memory footprint quickly decreases until $G$ = 16. It requires a bit more space at $G$ = 32 as the additional memory space for grouping matrices increases. Moreover, the breakdown chart on the right shows that the sparse row memory takes only 2.68\% of the total memory footprint due to its small entry number and the compact tuple format (bitvector: 512 bits, workload: 9 bits, maximum index: 4 bits). This is possible because we can only store a finite number of sparse data tuples in the sparse row memory, which is bounded by the group number, and reference it using the max indexes. As a result, OSEL compresses model parameters by 1.95$\times$ to 6.81$\times$ for different group numbers. By dramatically reducing the cycle count and memory space, LearningGroup enables efficient on-chip training without external memory access. 

\begin{table}
\footnotesize
\renewcommand{\arraystretch}{1.2}
\centering
\caption{Workload Deviation of Allocation Schemes from the Theoretical Workload }
\begin{tabular}{c|c|c|c|c} 
\hline
                                    & \textbf{G=2} & \textbf{G=4} & \textbf{G=8} & \textbf{G=16}  \\ 
\hline\hline
\textbf{Baseline (Threshold-based)} & 86.03        & 105.02       & 39.19        & 56.35          \\ 
\hline
\textbf{Proposed (Row-based)}       & 47.44        & 31.37        & 35.80        & 36.13          \\
\hline
\end{tabular}
\label{workload result}
\end{table}

\subsection{ Workload Allocation Efficiency}
To validate the LearningGroup's load balancing scheme, we track the workloads allocated to each core during the training of 2000 iterations. Table \ref{workload result} presents the maximum deviation from the theoretical workload, when the load balancing is perfectly applied. In the baseline, the threshold value is set by dividing the total number of unmasked elements in the weight matrix by the number of cores. It then distributes the unmasked elements to each core row-by-row until the number of assigned elements becomes larger than the threshold. On the other hand, the proposed row-based allocation evenly distributes the rows to cores, and this simpler allocation scheme constantly shows less variations than the threshold-based allocation. This is mainly because of the unaligned last workload assignments occurred in the threshold-based allocation. The row-based scheme works well because it distributes an equal number of rows to each core and the average sparsity of each core converges to $1/G$ over time. By using simple row-based allocation scheme, LearningGroup achieves 44.9\%, 70.1\%, 8.7\%, 35.9\% less deviations than the baseline when the group number is 2, 4, 8, and 16, respectively.
%achieves high hardware utilization by precisely balancing the load without additional hardware.

\begin{figure}[t!]
\centering
   \includegraphics[width=8.5cm, height=5.3cm]{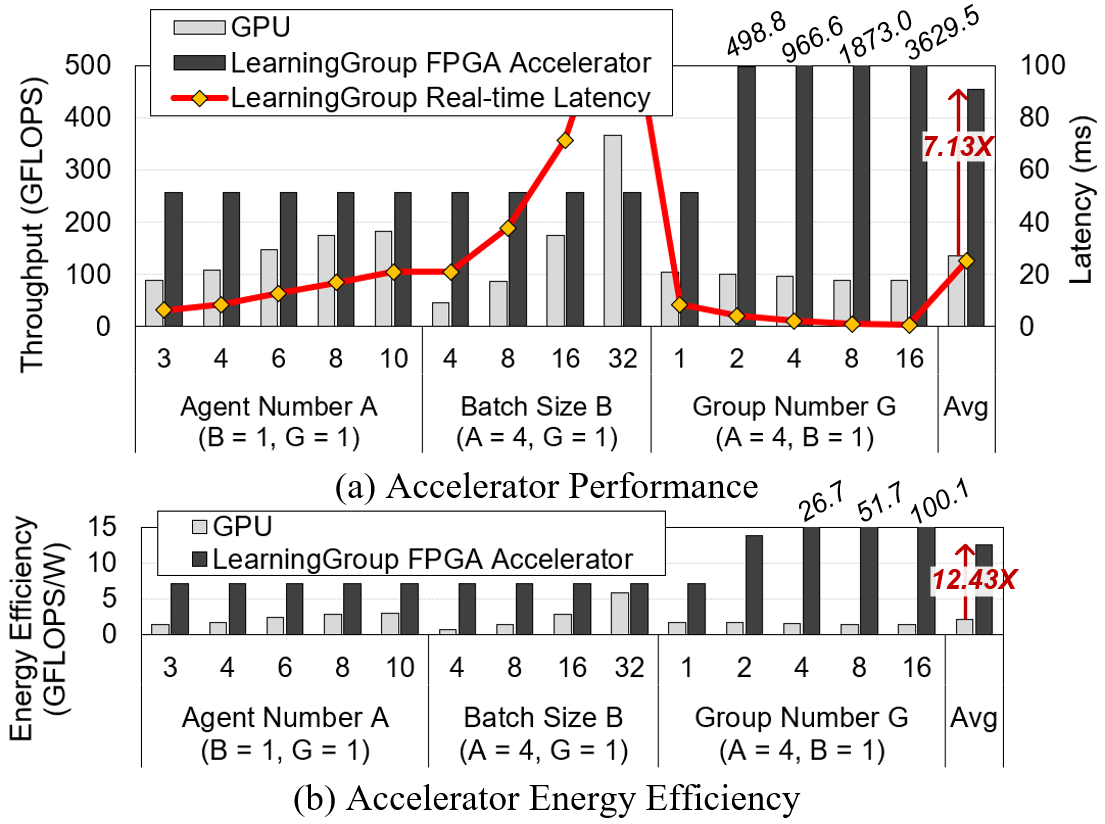}
\vspace{-0.1in}
\caption{Accelerator Performance Comparison}
\label{gpu comparison}
\vspace{-0.1in}
\end{figure}

\begin{figure}[t!]
\centering
   \includegraphics[width=8.5cm, height=2.5cm]{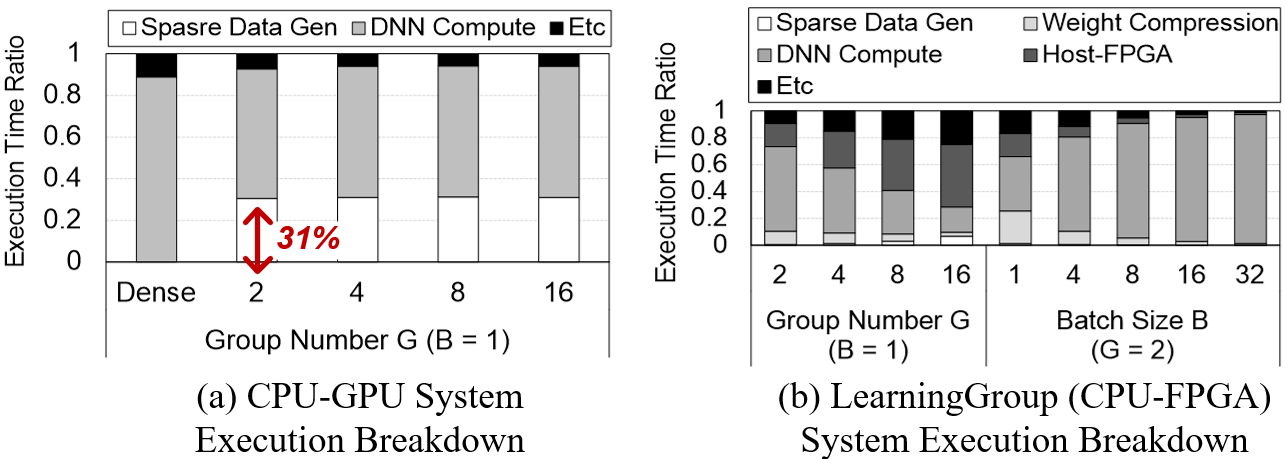}
\caption{System Execution Time Breakdown}
\label{system breakdown}
\end{figure}

\begin{figure*}[t!]
\centering
   \includegraphics[width=18cm, height=2.5cm]{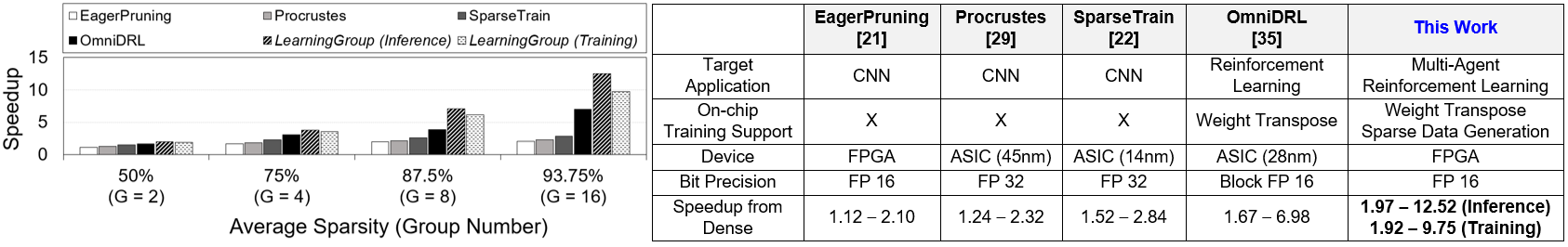}
\caption{Performance Comparison with Sparse Training Accelerators}
\label{sparse accelerator comparison}
\vspace{-0.2in}
\end{figure*}

\subsection{ Accelerator Performance}
\textbf{Comparison with GPU}
To evaluate the training performance, we compare the throughput and energy efficiency of LearningGroup's FPGA accelerator with the Nvidia Titan RTX GPU. With the same evaluation environment described in Section \ref{sec:eval-methodology}, our FPGA based system and the GPU show 36.3W and 63.18W power consumption on average for running the application, respectively. Figure \ref{gpu comparison}(a) and (b) show the throughput performance and energy efficiency measured only on the accelerators, i.e., FPGA and GPU. We exclude the host CPU time as it is same for both cases. We measure the accelerator performance under various scenarios: 1) varying the number of agents with a fixed batch size and group number, 2) varying the batch size with a fixed agent and group number, and 3) varying the group number with a fixed agent number and batch size. For the first and second scenarios, the throughput of our accelerator remains unchanged at 257.4 GFLOPS because it repeats the same computation for the agent number $A$ and for the batch size $B$ with a fixed hardware utilization. In the case of GPU, the throughput increases as the agent number and batch size increase because it has more workload to parallelize. Especially, it shows a linear performance gain with the batch size, but real-time MARL only allows small batch sizes less than 32 for faster training convergence and wide exploration.
\par
For the third scenario, GPU and our accelerator show completely opposite trends. By predicting the workload, our accelerator computes only non-zero operands with highly utilized VPUs, which enables a linear performance scaling as the group number increases. As a result, LearningGroup's FPGA accelerator achieves throughput performance up to 3629.5 GFLOPS as the group number is increased to 16, satisfying the computational requirement for real-time MARL processing.
In addition, LearningGroup shows an average latency of 25.04ms, smaller than the aforementioned real-time interaction constraint (30ms). If grouping is applied, it can have latency within 10ms and achieve high energy efficiency, unlike GPU. In the case of GPU, it does not benefit from the sparsity and remains at low throughput for all group numbers. This is because the grouping algorithm with sparse data generation is unsuitable for GPU. Figure \ref{system breakdown}(a) and (b) show the execution time breakdown of the GPU and our LearningGroup system when handling sparsity. We measure the ratio of the sparse data generation time and DNN computation time by changing the group number up to 16 with fixed batch size. In GPU, finding the maximum indexes in the grouping matrices and generating the mask matrix account for 31\% of the total execution time. GPU needs to fetch additional grouping matrices to generate the mask matrix, causing extra memory access. More importantly, GPU performs poorly on the masking operation required for accessing non-zero weights. This overhead can be hidden in the LearningGroup system. With OSEL, the sparse data generation accounts only for 2.9\% of the total execution time on average. This sparse data generation overhead can be further decreased as the batch size increases. Sparse data generation and weight compression are shared among the training batch samples, so the portion of DNN computation becomes dominant and attains the speedup through hardware-friendly generated sparsity. Thanks to the on-chip training and customization opportunity for irregular datapath/compact data format provided by FPGA, LearningGroup's FPGA accelerator achieves 454.80 GFLOPS throughput and 12.55 GFLOPS/W energy efficiency on average, which is 7.13$\times$ faster and 12.43$\times$ energy efficient than those of GPU, respectively.
\par
\textbf{Comparison with Sparse Training Accelerators}
For more evaluation, we compare LearningGroup against the state-of-the-art DNN training accelerators optimized for processing sparse data, such as EagerPruning \cite{zhang2019eager}, Procrustes \cite{yang2020procrustes}, SparseTrain \cite{dai2020sparsetrain}, and OmniDRL \cite{lee2021omnidrl}. Since each accelerator has a different number of processing units and operating frequency, we compare how well each accelerator exploits the sparsity by measuring the speedup achieved in sparse data over the dense case. Figure \ref{sparse accelerator comparison} shows the speedup performance of the accelerators with clarification in their hardware configurations. We measure the speedup of our accelerator for the sparse data generated by weight grouping, yielding 50\%, 75\%, 87.5\%, and 93.75\% sparsity. The speedup numbers of the other accelerators are calculated by interpolating their peak performances to the target sparsity. EagerPruning, Procrustes, and SparseTrain do not get a scalable gain with increasing sparsity because their gradual pruning algorithms are optimized for low sparsity. OmniDRL shows a notable gain over the other accelerators, but it still suffers from external memory access for the sparse training data. On the other hand, LearningGroup's speedup scales well at any sparsity level because it does not require any external memory access by performing both sparse data generation and weight transpose on-chip, enabling fast and energy efficient sparse training. LearningGroup itself shows a bit less speedup in training than in inference. The gap between the two becomes worse if the network gets sparser. This comes from the additional time to update the grouping matrices using the VPUs. Because the grouping matrix update occurs every iteration, like a normal weight update, it is not a significant bottleneck. As a result, LearningGroup achieves up to 12.52$\times$ and 9.75$\times$ speedup from the dense case for inference and training, respectively, which are the best among the state-of-the-art works by showing 5.96$\times$ and 4.64$\times$ improvement than EagerPruning.

\section{ Related Work} \label{sec:related}
{\bf Deep Reinforcement Learning Accelerator}
There have been an increasing number of works on accelerating deep reinforcement learning while they only focus on single-agent domain \cite{cho2019fa3c, kim20192, lee2021omnidrl, yang2021fixar}. Kim et al. \cite{kim20192} compress the input activations using the top three frequently used exponent values rather than pruning the weights. OmniDRL \cite{lee2021omnidrl} also uses weight pruning, but its pruning algorithm degrades the reward when applied to MARL. LearningGroup is the first work that targets online multi-agent reinforcement learning by utilizing a hardware-amenable pruning algorithm. 
\par
{\bf Sparsity Handling Accelerator}
EagerPruning \cite{zhang2019eager} is the first sparse training accelerator that gradually prunes the lowest-magnitude weights. It is not hardware-amenable in terms of latency and energy by relying on the sorting algorithm. Procrustes \cite{yang2020procrustes} improves the sparse training efficiency using a load-balanced dataflow while it still suffers a low compression ratio. All other sparse accelerator targets only inference or extremely high sparsity. Many works \cite{pal2018outerspace, zhang2020sparch, parashar2017scnn, gondimalla2019sparten} use variants of compressed sparse row/column format (CSR/CSC). However, when the sparsity is less than 90\%, the proposed bitvector based format shows a higher compression ratio than CSR/CSC with easier address calculation \cite{dave2021hardware}. Since most model pruning algorithms settle in the sparsity level under 90\% due to the accuracy issue, LearningGroup can be used for other DNN inference/training workloads.
\section{ Conclusion} \label{sec:conclude}
In this paper, we present an FPGA-based multi-agent reinforcement learning acceleration system called LearningGroup, which applies network pruning on the training of MARL for the first time. We propose an on-chip spare data encoding loop (OSEL) that enables fast encoding and compression for transposed data. We devise a simple but effective row-based load balancing scheme at run-time while taking intra-layer parallelism together. Finally, we design a core with vector processing units that process multiple compressed weight matrix rows with high utilization based on load balancing. As a result, we reduce the cycle and memory space for sparse data generation up to 5.72$\times$ and 6.81$\times$, respectively. Our accelerator shows 7.13$\times$ speedup and 12.43$\times$ energy efficiency on average than those of GPU, respectively, by supporting fully on-chip training and highly optimized dataflow/data format based on FPGA's unique advantage. Most significantly, it achieves 12.52$\times$ speedup when processing sparse data over the dense case, which is the highest speedup compared to the previous sparse training accelerator, thanks to the algorithm/architecture co-design approach.
\section*{ACKNOWLEDGMENT}
This research was supported in part by the Information Technology Research Center (ITRC) support program (IITP-2020-0-01847) and the grant (No.2022-0-01036, Development of Ultra-Performance PIM Processor SoC with PFLOPS-Performance and GByte-Memory), both supervised by the Institute for Information \& Communications Technology Planning \& Evaluation (IITP) under the Ministry of Science and ICT (MSIT), Republic of Korea.

%%%%%%%%% -- BIB STYLE AND FILE -- %%%%%%%%
\begingroup
\setstretch{0.89}
\bibliographystyle{IEEEtran}
\bibliography{IEEEabrv, 8_reference}

\endgroup
%%%%%%%%%%%%%%%%%%%%%%%%%%%%%%%%%%%%

\end{document}